\documentclass[pra,unsortedaddress,showpacs,preprint]{revtex4-1}
\usepackage{epsfig}
\usepackage{graphicx}
\usepackage{amsfonts}
\usepackage{color}
\usepackage{multirow}

\begin{document}

\title{ Ground-state stability and  criticality  of 
two-electron atoms with screened Coulomb potentials using the  B-splines basis set}

\author{Pablo Serra}
\email{ serra@famaf.unc.edu.ar}
\affiliation{Department of Chemistry, Purdue University,
West Lafayette, Indiana 47907, USA}

\affiliation{Facultad de Matem\'atica, Astronom\'{\i}a y F\'{\i}sica,
Universidad Nacional de C\'ordoba, and IFEG-CONICET, Ciudad Universitaria,
X5000HUA,  C\'ordoba, Argentina}

\author{Sabre Kais}
\email{kais@purdue.edu}

\affiliation{Department of Chemistry, Purdue University,
West Lafayette, Indiana 47907, USA}

\affiliation{Department of  Physics,  Purdue University, 
West Lafayette, Indiana 47907, USA}

\affiliation{Qatar Environment and Energy Research Institute, Qatar Foundation,
Doha, Qatar}

\begin{abstract}
{We applied  the finite-size scaling method using the B-splines basis set  to construct the stability diagram for two-electron atoms 
 with  a screened  Coulomb potential.   The results of this  method  for two electron atoms are very accurate in comparison with previous calculations  based on 
Gaussian, Hylleraas, and finite-element basis sets. The stability diagram for the screened two-electron 
atoms shows three distinct regions: a two-electron region, a one-electron region,
and a zero-electron region, which correspond to stable, ionized and double ionized atoms. In previous studies, it was difficult to extend the finite size scaling calculations to large molecules and extended systems because of the computational cost and the 
lack of a simple way to increase the number of Gaussian basis elements in a systematic
 way. Motivated by recent studies showing how one can use B-splines to solve 
Hartree-Fock and Kohn-Sham equations, this combined finite size scaling using the B-splines basis set, might provide an effective systematic way to treat criticality of large molecules and extended systems. As benchmark calculations, the two-electron systems show the feasibility of this combined approach and provide an accurate reference for comparison.   
}
\end{abstract}

\pacs{}
\maketitle
\section{Introduction}

Weakly bound systems represent an interesting field of research in atomic and molecular physics. 
The behavior
of systems near a binding threshold is important in the study of ionization of atoms and molecules, molecule
dissociation, and scattering collisions.
Since the pioneering works of  Bethe \cite{bethe} and Hylleraas\cite{hylleraas} confirming the existence of
 the  negative hydrogen ion, $H^-$, the study of the stability of the ground state of atomic and molecular negative anions becomes  an active 
field of research. New phenomena appear when the Coulomb interaction is screened  and the long-range 
electrostatic interactions turn to  short-range potentials. A simple model to describe the effect of the 
screening in the  Coulomb potential is the Yukawa potential, where an exponential decay is introduced,
$1/r\;\rightarrow  \;\exp{(-r/D)}/r$, where $D$ is a positive constant. The Yukawa potential has been 
used in many branches of physics, for example to describe interactions in dusty plasmas
where charged dust particles are surrounded by plasma \cite{s98},
liquid metals \cite{hcr96}, charged  colloidal particles\cite{hhbn11}.  Two-electron systems interacting via Yukawa potentials were the subject of  recent research, 
studying the bound states using the Hylleraas basis set \cite{sh06,ps09,khj11,ssm12,jkh12} 
and B-spline  expansions \cite{llh12}. 
Also,  scattering processes were recently presented \cite{zfbj11} using  Yukawa potentials.

To examine near threshold  behavior, the  Finite Size Scaling (FSS) approach is
needed in order to extrapolate results obtained  from finite systems to the complete basis set limit.  
FSS is not only a formal way to understand
the asymptotic behavior of a system when the size tends to infinity, but a
theory that also gives us numerical
methods capable of obtaining
accurate results for infinite systems by studying the corresponding small
systems\cite{neirotti0,serra2,kkk,snk1,snk2,nsk,qicun,kais1,review,adv,dipole,quadrupole}. 
Applications  include expansion in Slater-type basis functions\cite{adv},  
Gaussian-type  basis functions\cite{gto} and recently, 
finite elements\cite{fem-fss}.

Here, we combine FSS with the B-splines expansion to calculate the  stability diagram for two-electron atoms 
 with  a screened  Coulomb potential. The B-spline
functions $B_i(r), i=1,n_s$   form a basis for piecewise polynomial functions that are polynomials of degree $(k_s  -  1)$  in each interval and whose
derivatives up to order $(k_s   -  2)$  are continuous at the interior knots, have been increasingly used in atomic and molecular physics\cite{B-s1,B-s2,B-s3,B-s4,deboor}.
Our results show the B-splines functions are very efficient in performing FSS to calculate the critical parameters and the stability diagram.

The paper is organized as follows. In Section \ref{sec-two} we present FSS with a B-splines basis
followed by the two-electron atom, as a benchmark calculation in Section \ref{sec-three}.
In Section~\ref{sec-four}, we present our main results for the screened two-electron atoms. Finally, 
in Section~\ref{sec-conclu} we discuss our results and conclusions.

\section{Finite size scaling (FSS) with B-splines}
\label{sec-two}

Here, we briefly introduce the finite size scaling (for more details, see Ref. \cite{adv}) and how 
to perform calculations using B-splines. 
The finite size scaling method is a systematic way to extract
the critical behavior of an infinite system from analysis on finite
systems \cite{adv}. It is efficient and accurate for the calculation of
critical parameters for few-body  Schr\"odinger equation.  

In our study, we have Hamiltonian of the 
 following form:
\begin{equation}
\label{h1}
{\cal H} \,=\, {\cal H}_0 \,+\, V_\lambda \;
\end{equation}

\noindent where ${\cal H}_0$ is $\lambda$-independent and $ V_\lambda$
is the $\lambda$-dependent  term. We are interested in the study of
how the different properties of the system change when the value of
$\lambda$ varies.
A critical point, $\lambda_c$,  will be  defined as a point for which a
bound state becomes absorbed or degenerate with a continuum.  We also define a critical exponent 
$\alpha$ by the asymptotic behavior of the ionization energy $E(\lambda)-{\cal E}_{th}\sim 
(\lambda-\lambda_c)^\alpha$, where we assume that the threshold energy, ${\cal E}_{th}$, does not 
depend on $\lambda$.
 In the first example, the He-like atoms, we have only one parameter, $\lambda$
while for the second case, the screened two electron atoms, we have two parameters, $\lambda_1$ and $\lambda_2$.  
To perform the finite size scaling calculations, we expand the exact wave  function in a finite basis set and  truncate this expansion at some
order $N$. The finite
size corresponds to the number of elements
in a complete basis set used to expand the exact  eigenfunction
of a given Hamiltonian. The ground-state eigenfunction has
the following expansion:$ \Psi_\lambda= \sum_{n}^{}a_n(\lambda)\phi_n$,
where n is the set of quantum numbers. We have to truncate the
series at order N, and the expectation value of any general operator ${\cal O}$ at
order N is given by:
\begin{equation}
\left<{\cal O}\right>^{N} = \sum_{n,m}^N a_n^{(N)}a_m^{(N)} {\cal O}_{n,m},
\end{equation}
\noindent where ${\cal O}_{n,m}$ are the matrix elements of ${\cal O}$
in the basis set $\{\phi_n\}$.

In this study we used the B-splines basis,  the  normalized one-electron orbitals are given by
\begin{equation}\label{phi-bs}
\phi_{n}({r}) = C_n \, \frac{B^{(k)}_{n+1}(r)}{r}  \,;\;\;n=1,\ldots
\end{equation}

\noindent where $B^{(k)}_{n+1}(r)$ is a B-splines polynomial of order $k$.
The numerical results  are obtained by defining a cutoff radius $R$, and
then the interval $[0,R]$ is divided into $I$ equal subintervals. 
 B-spline polynomials \cite{deboor} (for a review
of applications of B-splines polynomials in atomic and molecular physics,
see ref. \cite{bachau01})  are piecewise polynomials defined by a 
sequence of  knots $t_1=0\leq t_2\leq\cdots \leq t_{2 k+I-1}=R$ 
and the recurrence relations

\begin{equation}\label{bs1}
B_{i,1}(r)\,=\,\left\{ \begin{array}{ll} 1 & \mbox{if}\,t_i\leq r < t_{i+1}   \\
0 &\mbox{otherwise,}  \end{array}  \right. \,.
\end{equation}

\begin{equation}\label{bsrr}
B_{i,k}(r)\,=\,\frac{r-t_i}{t_{i+k-1}-t_i}\,B_{i,k-1}(r)\,+\,
\frac{t_{i+k}-r}{t_{i+k}-t_{i+1}}\,B_{i,k-1}(r)\; (k>1)\,.
\end{equation}

\noindent In this work, we use the standard  choice for the knots  in 
atomic physics  \cite{bachau01} $t_1=\cdots=t_k=0$ and $t_{k+I}=\cdots=t_{2k+I-1}=R$.
Because we are interested in FSS, we choose an equidistant distribution of
inside knots. The constant $C_n$ in Eq.(3) is a normalization 
constant obtained from the condition $\langle n | n \rangle=1$,

\begin{equation}\label{nor-c}
C_n = \frac{1}{\left[ \int_0^{R_0} \, \left(B^{(k)}_{n+1}(r) \right)^2 \,dr
\right]^{1/2}} \,.
\end{equation}

Because $B_1(0)\ne0$ and $B_{I+k-1}(R)\ne0$, we have $N=I+k-3$ orbitals 
corresponding to $B_2,\ldots,B_{I+k-2}$. In all the calculations we used the value $k=5$, and,
 we do not write the index $k$ in the eigenvalues and coefficients.

To obtain the numerical values of the critical parameters $(\lambda_c, \alpha)$ for the  energy,
we define for any given operator ${\cal O}$ the function

\begin{equation}
\triangle_{\cal O}(\lambda;N,N')=\frac{\ln\left(\left<{\cal O}_\lambda^{N} \right>/
\left<{\cal O}\right>_\lambda^{N'}\right)} {\ln(N'/N)},
\label{fourteen}
\end{equation}

If we take the operator ${\cal O}$ to be $H-{\cal E}_{th}$, and  $\partial H/\partial \lambda$,
we can obtain the critical  parameters from the following function \cite{adv}
 
\begin{equation}
\Gamma_\alpha(\lambda,N,N')=\frac{\triangle_H(\lambda;N,N')}{\triangle_H(\lambda;N,N')-\triangle_{\frac{\partial
      H}{\partial \lambda}}(\lambda;N,N')},
\label{gammafunc}
\end{equation}
which at the critical point is independent of $N$ and $N'$ and takes
the value of $\alpha$. Namely, for $\lambda=\lambda_c$ and any
values of $N$ and $N'$ we have
\begin{equation}
\Gamma_\alpha(\lambda_c,N,N')=\alpha.
\end{equation}

\noindent Because our results are asymptotic for large values of $N$, we obtain a sequence of 
pseudocritical parameters $(\lambda_N,\,\alpha_N)$ that converge to $(\lambda_c,\,\alpha)$
for $N \rightarrow \infty$.


\section{Helium-like atoms}
\label{sec-three}

As a benchmark for FSS using B-splines we calculate the critical parameters
of the two-electron atom with standard Coulomb potential.  In this case, after
a scaling with the nuclear charge, 
the system has a unique parameter $\lambda=1/Z$

\begin{equation}
\label{hamilhe}
H = -\frac{1}{2} \nabla_{{\mathbf r}_1}^2
-\frac{1}{2} \nabla_{{\mathbf r}_2}^2 \,-\,
\frac{1}{r_1}\,-\,\frac{1}{r_2}\,+\,\lambda \,
\frac{1}{\left|{\mathbf r}_2-{\mathbf r}_1\right|} \,.
\end{equation}

The ground-state energy $E_0(\lambda_1,\lambda_2)$ and its corresponding
eigenvector $\left|\psi_0(1,2)\right\rangle $ will be calculated within the
variational approximation

\begin{equation}\label{variational-functions}
\left|\psi_0(1,2)\right\rangle \,  \simeq\,
\left|\Psi_0(1,2)\right\rangle \, =\,
\sum_{i=1}^M c^{(j)}_{i} \left| \Phi_i
\right\rangle \, ,\;\; c^{(j)}_{i} = (\mathbf{c}^{(j)})_i
\;\;;\;\;j=1,\cdots,M \,.
\end{equation}

\noindent where the $\left| \Phi_i \right\rangle$ must be chosen adequately and $M$ is the
 basis set size.

Since we are interested in the behavior of the system near the
ground-state ionization
threshold, we choose as basis set  s-wave singlets given by

\begin{equation}\label{basis}
\left| \Phi_i\right\rangle \equiv \left| n_1,n_2;l\right\rangle =
\left( \phi_{n_1}({r}_1) \,
\phi_{n_2}({r}_2)  \right)_s
\mathcal{Y}_{0,0}^l (\Omega_1,\Omega_2) \, \chi_{s} \, ,
\end{equation}
where $n_2\leq n_1\leq N$. Also, we introduce  a cutoff value $l_{max}$ for the
angular
momentum  $l\leq l_{max}$, denote $\chi_{s}$ as the singlet spinor,
 and the $\mathcal{Y}_{0,0}^l (\Omega_1,\Omega_2) $ are given by
\begin{equation}\label{angular-2e}
\mathcal{Y}_{0,0}^l (\Omega_1,\Omega_2)\,=\, \frac{(-1)^l}{\sqrt{2l+1}} \,
\sum_{m=-l}^{l} (-1)^m Y_{l\,m}(\Omega_1) Y_{l\, -m}(\Omega_2) \, ,
\end{equation}
{\em i.e.} they are eigenfunctions of the total angular momentum with zero
eigenvalue, and
the $Y_{l\, m}$ are the spherical harmonics. {  Note also that $\mathcal{Y}_{0,0}^l$
 is a real function since it is symmetric in the particle index.}  The  radial term
$(\phi_{n_1}({r}_1) \phi_{n_2}({r}_2))_s$
has the appropriate symmetry for a singlet state,

\begin{equation}\label{radial-sym}
(\phi_{n_1}({r}_1) \phi_{n_2}({r}_2))_s \,=\, \phi_{n_1}(r_1)
\phi_{n_2}(r_2)+ \phi_{n_1}(r_2) \phi_{n_2}(r_1) \,.
\end{equation}
In general, the size $M$ of a basis set defined for Eqs.(\ref{variational-functions}-\ref{radial-sym}) 
is $M=N (N+1)(l_{max}+1)/2$.
For the    radial orbitals we used normalized B-splines polynomial
of order $k$
\begin{equation}\label{phi-bs2}
\phi_{n}({r}) = C_n \, \frac{B^{(k)}_{n+1}(r)}{r}  \,;\;\;n=1,\ldots,N=k+I-3 \,.
\end{equation}

\noindent The  calculations in this sections were done with $k=5,\,R=30$, and $l_{max}=3$.

In order to calculate the Hamiltonian matrix elements, we  expand
 the electronic Coulomb interaction in spherical harmonics

\begin{equation}
\frac{1}{\left|{\mathbf r}_2-{\mathbf r}_1\right|}\,=\,
\sum_{l=0}^\infty \,\frac{4 \pi}{2 l+1} \,
\frac{r_<^l}{r_>^{l+1}}\,\sum_{m=-l}^l \,Y_{l,m}^*(\Omega_1)\,Y_{l,m}(\Omega_2),
\end{equation}

Because the cutoff  $l_{max}$, the  matrix elements of this expansion are nonzero
 only for $l\leq 2 l_{max}$.

In our previous studies, the critical behavior of the two-electron atom was obtained by using   FSS
approach with  Hylleraas \cite{nsk98} and Gaussian basis sets \cite{mks08}. 
The FSS  was  performed  with a finite small  basis-set and then  increased the number of basis 
functions $N$ in  a systematic way.   The B-splines
basis-set, in this sense is different. When $N$ is changed, we are not adding
new functions, but the complete basis-set is changed in a way which is  similar to 
the  finite-element method \cite{awk12}.

Figure 1, shows  the results for the plot $\Gamma_N$ as  $\lambda$  varies for different values of $N$ . Successive  curves 
cross  at pseudocritical  points. In Figures  (2) and (3), we observed the behavior of the pseudocritical 
parameters, 
$\lambda_N$ and $\alpha_N$ as a function of $1/N$. The two curves converged to the 
exact values, in complete agreement with our previous \cite{snk2,gto,adv} and
recent  results\cite{gt11}. The numerical values are shown in table \ref{table_comp}. 
These accurate results indicate that FSS can be combined with B-spline basis to obtain critical parameters 
for  the few-body Schr\"odinger equation.

\begin{figure}[floatfix]
\begin{center}
\psfig{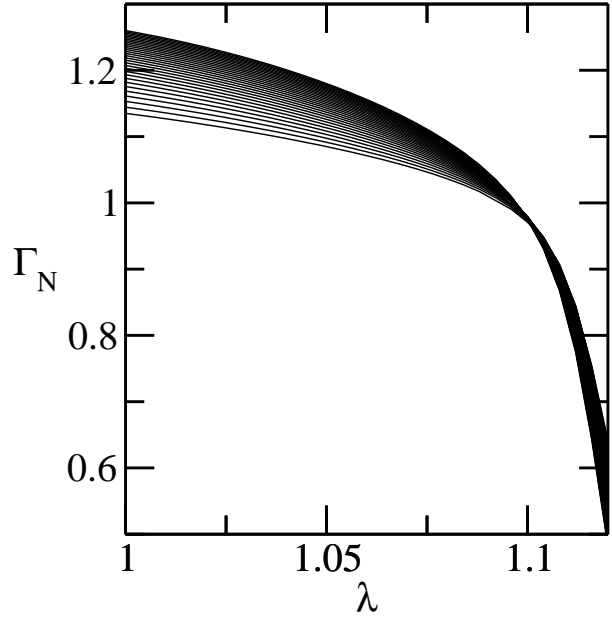}
\end{center}
\caption{\label{ygamma-He}(color-online)  $\Gamma_N$ vs. $\lambda$  
for two-electron  atoms, for $N=20,\cdots,50$.}
\end{figure}

\begin{figure}[floatfix]
\begin{center}
\psfig{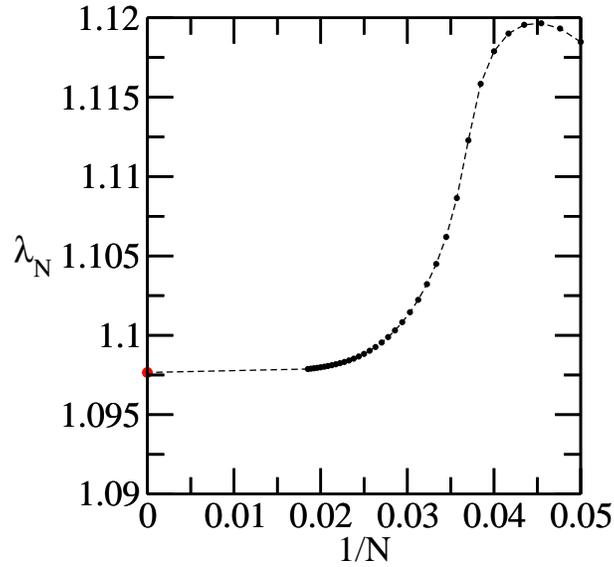}
\end{center}
\caption{\label{he-lc} $\lambda_N$ vs. $1/N$  for $N=20,\cdots,55$ for the two-electron atom. The
red point is the value of $\lambda_c$ from ref.\cite{nsk98} .}
\end{figure}

\begin{figure}[floatfix]
\begin{center}
\psfig{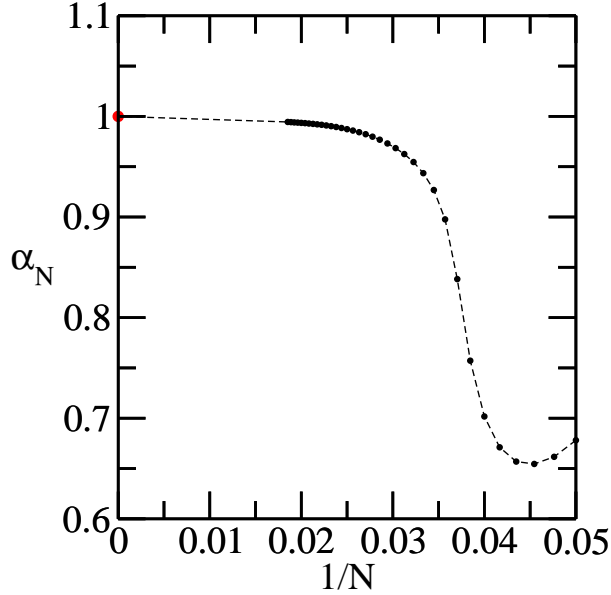}
\end{center}
\caption{\label{he_alpha}(color-online)  Critical exponent $\alpha_N$ vs. $1/N$
 for $N=20,\cdots,55$  for the two-electron atom. The red point is the exact value $\alpha =1$.}
\end{figure}

\begin{table*}[th!]
\begin{tabular}{|c|c|c|c|c|c|c|c}\hline\hline
\hspace{.1cm}    \hspace{.1cm}   &
\hspace{.2cm}  This work  \hspace{.2cm}   &
\hspace{.2cm}  FSS (Hylleraas) \cite{adv}  \hspace{.2cm}   &
\hspace{.2cm}  Ref \cite{gt11}  \hspace{.2cm}   &
\hspace{.2cm}  exact  \hspace{.2cm}   \\ \hline \hline 
$\lambda_c$        &   1.09776   & $1.0976$   &  $1.09788$   &  $-$  \\
\hline
$ \alpha $        & $0.9947$          & $1.04$          & $-$  & $1$ \\
\hline \hline
\end{tabular}
\caption{\label{table_comp} Comparison of $\lambda_c $ and 
$\alpha $ for the ground-state energy of the two-electron
atom.}
\end{table*}



\section{The screened two-electron atom}
\label{sec-four}

The Hamiltonian, in atomic units, takes the form,
\begin{equation}
\label{hamiltoniano}
H = -\frac{1}{2} \nabla_{{\mathbf r}_1}^2  
-\frac{1}{2} \nabla_{{\mathbf r}_2}^2  \,-\,
\frac{Z\,e^{-r_1/D}}{r_1}\,-\,\frac{Z\,e^{-r_2/D}}{r_2}\,+\,
\frac{e^{-\left|{\mathbf r}_2-{\mathbf r}_1\right|/D}}{
\left|{\mathbf r}_2-{\mathbf r}_1\right|} ,
\end{equation}

\noindent where $Z$ is the nuclear charge and $D>0$ the Debye screening
length. The Hamiltonian takes a form more convenient to our purposes after
scaling with $D$, $r \rightarrow r/D$ and $H  \rightarrow D^2 \,H \,$

\begin{equation}
\label{hamil}
H = -\frac{1}{2} \nabla_{{\mathbf r}_1}^2  
-\frac{1}{2} \nabla_{{\mathbf r}_2}^2 \,-\,\lambda_1\,\left(
\frac{e^{-r_1}}{r_1}\,+\,\frac{e^{-r_2}}{r_2}\right)\,+\,\lambda_2\,
\frac{e^{-\left|{\mathbf r}_2-{\mathbf r}_1\right|}}{
\left|{\mathbf r}_2-{\mathbf r}_1\right|} \,,
\end{equation}

\noindent where $\lambda_1=Z\,D$, and $\lambda_2=D$.
The numerical results  are obtained using the same basis set as the Coulomb case described in section
III, except the value of the  cutoff radius, that for the Yukawa potential we set as  $R=20$.

For the Yukawa potential we  use the Gegenbauer's expansion in 
spherical harmonics

\begin{equation}
\frac{e^{-\left|{\mathbf r}_2-{\mathbf r}_1\right|}}{\left|{\mathbf r}_2-
{\mathbf r}_1\right|}\,=\,\sum_{l=0}^\infty \,4 \pi \,
\frac{I_{l+1/2}(r_<)}{\sqrt{r_<}}\,
\frac{K_{l+1/2}(r_>)}{\sqrt{r_>}}\,
\,\sum_{m=-l}^l \,Y_{l,m}^*(\Omega_1)\,Y_{l,m}(\Omega_2),
\end{equation}

\noindent where $I_{l+1/2}$ and $K_{l+1/2}$ are the modified Bessel functions 
of the first and second kind, respectively \cite{abramowitz}.

\begin{figure}[floatfix]
\begin{center}
\psfig{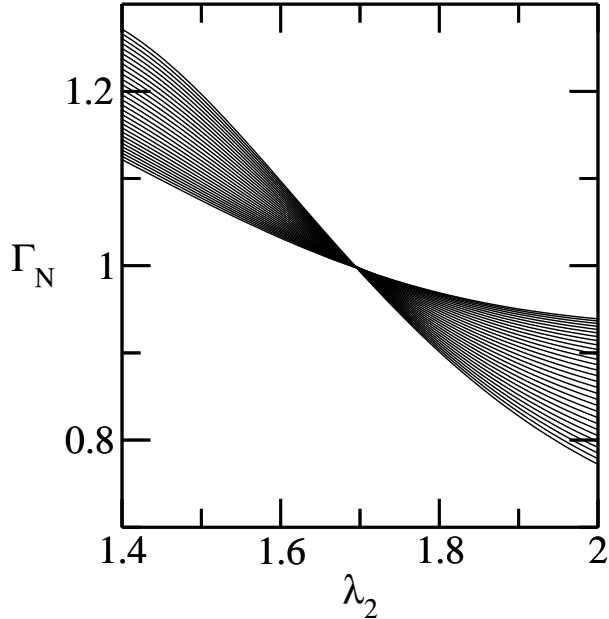}
\end{center}
\caption{\label{ygamma}(color-online)  $\Gamma_N(\lambda_1=1.5;\lambda_2)$ vs. $\lambda_2$  
for the screened two-electron  atom, for $N=20,\cdots,50$.}
\end{figure}

\begin{figure}[floatfix]
\begin{center}
\psfig{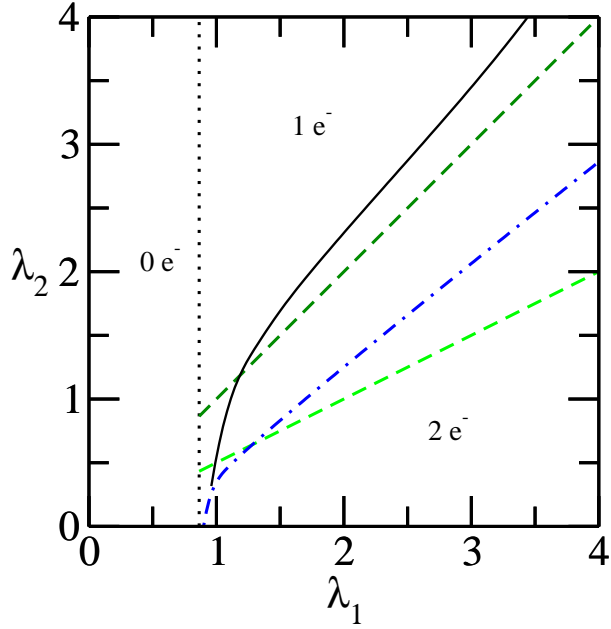}
\end{center}
\caption{\label{ypd}(color-online)  Ground-state stability diagram for the screened two-electron
 atom, the black line is  the critical $2e^--1e^-$ line calculated with FSS with  $N=40$, 
 the dot line is the $1e^--0e^-$ critical line, the dashed
blue line is the  lower bound for the  $2e^--1e^-$ line of ref. \cite{ps08}. 
The  dot-dashed lines correspond
to the Helium (dark green) and Hydrogen (light green) atoms respectively.}
\end{figure}

Since ${\cal E}_{th}$ does not depend on $\lambda_2$, we calculate the scaling function 
$\Gamma_N$ for given  values of $\lambda_1$ as a function of $\lambda_2$.
Figure \ref{ygamma}, show the results for the plot   $\Gamma_N(\lambda_1=1.5;\lambda_2)$  as  $\lambda_2$ varies
for different values if $N$. All the curves
cross very close to the critical point.   In Figure  \ref{ypd} we present the phase diagram for the screened two-electron atoms with
 three distinct phases: two-electron phase ($2e^-$), one-electron  phase ($1e^-$)
and zero-electron phase ($0e^-$), corresponding to stable, ionized and double ionized atoms.

The dotted line $\lambda_1=\lambda_1^{(c)}\simeq 0.84$ corresponds to the critical value of the 
one-electron Yukawa potential. Therefore, there are no bound states for $\lambda_1
\leq\lambda_1^{(c)}$. For $\lambda_1 \gtrsim \lambda_1^{(c)}$ the one-body bound state is  extended,
and the method is applicable until the size of the one-body  state 
becomes of the  order of the cutoff radius $R$. For the value $R=20$, we calculate the $1e^-$stability 
line for  $\lambda_1 \ge0.95$.

Reference \cite{ps08}  described the three different  ground-state stability diagrams that a
two-parameter Hamiltonian with short-range one-body potential could present. These cases are 
(see figure 1 of this reference)
(a) no $2e^--0e^-$ line,  (b) exists a  finite  $2e^--0e^-$ line for 
$0\leq \lambda_2\leq\lambda_2^{(mc)}$, (c) the $2e^--0e^-$ line is infinite. 
Also in this reference rigorous lower
and upper bound for the $2e^--1e^-$ stability line are established. We calculate these bounds for the 
Hamiltonian Eq(\ref{hamil}). The lower bound is shown in figure \ref{ypd}. For the
Yukawa potential  the upper
bound diverges for $\lambda_1 \rightarrow\lambda_1^{(c)}$, and then it is not useful in this case.

Even our results suggest that the ground-state stability diagram  is of type (a). Large numerical 
instabilities could appear for  $\lambda_1 \rightarrow \lambda_1^{(c)}$, and then we can discard a
type (c) diagram, but we can not  discard  a type (b) diagram with a small value of $\lambda_2^{(mc)}$.

We note that the $H^-$ atom corresponds to the line $\lambda_2=\lambda_1$ and the He atom to the line
$\lambda_2=\lambda_1/2$. These lines are also indicated in figure  \ref{ypd}. The critical 
screening values for H$^-$ and He are $D_{H^-}\simeq 1.2969 $ and $D_{He}\simeq 0.4934 $ respectively.

\begin{figure}[floatfix]
\begin{center}
\psfig{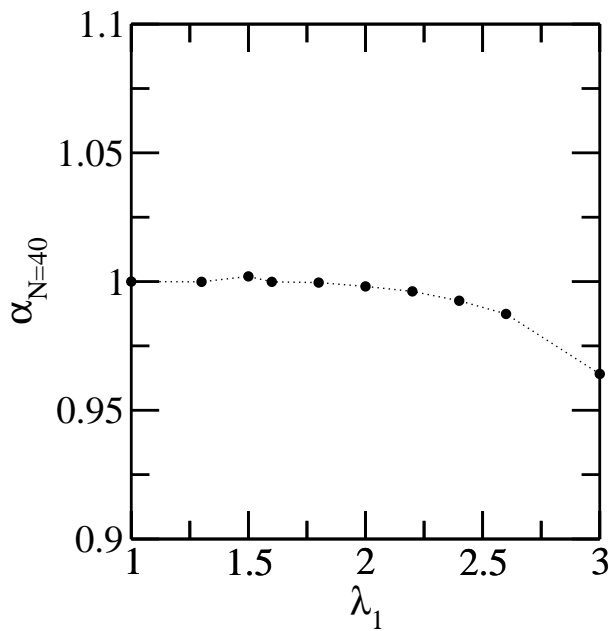}
\end{center}
\caption{\label{yalpha}(color-online)  Critical exponent $\alpha_N$ vs $\lambda_1$ for $N=40$
for the  screened two-electron
 atom. The exact value is $\alpha=1$.}
\end{figure}

In Figure \ref{yalpha} we show how the  critical exponent $\alpha^N$ vs $\lambda_1$ for $N=40$
for the  screened two-electron atom converges to the  exact value,  $\alpha=1$ \cite{ps08}.

\section{summary and conclusions}
\label{sec-conclu}

We have shown that the  introduction of B-spline basis sets in FSS  calculations is very powerful in 
obtaining critical parameters and stability diagrams for
few-body systems.

This basis set presents very different  characteristics than the standard basis sets previously used
in FSS like Hylleraas or Slater-type basis sets. B-splines are non-zero only on a small interval, 
and changing the FSS parameter N (number of basis functions) changes the complete basis set.
In particular, we used  this basis set together with FSS to calculate the critical 
parameter of the helium-like atom as a benchmark, finding very accurate results. 
We then applied the method to the 
important case of the ground-state stability diagram for a two-electron atom  
interacting via  a screened 
Coulomb potential.  Also in this case, FSS with a B-spline basis-set proves to be an excellent approach 
to obtain the critical behavior  for this two-parameter Hamiltonian.

Our results  show that the ground-state  diagram of two-electron atoms interacting via Yukawa 
potentials does not present a $2e^--0e^-$ line. That is, 
the systems always undergoes  a $2e^--1e^-$ transition before losing both electrons as the 
screening grows. Even the  numerical results are not accurate enough  to discard a small $2e^--0e^-$ line. 
We discard  the  existence of an infinite $2e^--0e^-$ line.

We have shown in previous works that  FSS  combined with different basis functions
(Hylleraas, Gaussian, Slater) is a powerful method to  obtain quantum critical parameters for 
few-body systems \cite{adv}. However,  these basis sets are not useful to calculate critical 
parameter for large systems, or for quantum phase transitions in infinite systems.
A possible way to apply FSS to study quantum phase transitions in materials is to combine FSS 
with Hatree-Fock or density functional approaches. In this direction,  new efficient methods to 
solve  the Hartree-Fock equations using B-splines expansions were recently established 
 \cite{ff07}, and numerical codes are available \cite{B-s1}. 
 As a benchmark system, we started with the two electron atoms.  
We show that indeed this can be done and obtained very accurate quantum critical 
parameters.  Then we went to a more difficult case,  two-electron atoms with screened
 Coulomb potentials.  Getting all the stability and transition lines from 
two-electrons to one-electron to zero-electrons is numerically difficult calculations.
 We have shown that FSS with B-spline basis functions can construct the full stability
 diagram. Our work is in progress to calculate critical parameters for large i
molecular and extended  systems by  applying FSS with  B-spline expansions of 
Hatree-Fock equations.

\acknowledgments
This work is supported by the NSF Centers
for Chemical Innovation: Quantum Information
and Computation for  Chemistry,
CHE-1037992.   P. Serra  would like to acknowledge the hospitality of Purdue University, where the work was done, and partial financial support of  SECYT-UNC,  CONICET, 
Min-CyT-C\'ordoba and the {\em Programa Cuarto Centenario de la Universidad 
Nacional de C\'ordoba}.

\end{document}